\documentclass[12pt,preprint]{aastex}
\slugcomment{}

\shortauthors{H. Tsunemi et al.}
\shorttitle{Improvement of the Spatial Resolution of the ACIS Using Split Pixel Events}
\newcommand{\AGNa}{PKS0312$-$770}
\newcommand{\AGNb}{PG1634+706}
\newcommand{\AGNL}{Q0957+561}

\begin{document}

\title{Improvement of the Spatial Resolution of the ACIS Using Split Pixel Events}

\author{Hiroshi Tsunemi\altaffilmark{1,2},
Koji Mori\altaffilmark{1, 3},
Emi Miyata\altaffilmark{1,2},
Christopher Baluta\altaffilmark{1},
David N. Burrows\altaffilmark{3},
Gordon P. Garmire\altaffilmark{3},
and
George Chartas\altaffilmark{3}
}

\altaffiltext{1}{Department of Earth and Space Science,
	Graduate School of Science, Osaka University,
	1-1 Machikaneyama, Toyonaka, Osaka 560-0043 JAPAN;
	tsunemi@ess.sci.osaka-u.ac.jp, mori@ess.sci.osaka-u.ac.jp,
	miyata@ess.sci.osaka-u.ac.jp, baluta@ess.sci.osaka-u.ac.jp}
\altaffiltext{2}{CREST, Japan Science and Technology Corporation (JST)}
\altaffiltext{3}{Department of Astronomy \& Astrophysics, 525 Davey
Laboratory, Penn State University, University Park, PA 16802, U.S.A.;
burrows@astro.psu.edu, garmire@acis.astro.psu.edu,
chartas@lonestar.astro.psu.edu}

\begin{abstract}

The position accuracy of X-ray photons on a CCD detector is generally
believed to be limited by the CCD pixel size.  While this is true in
general, the position accuracy for X-ray events which deposit charge in
more than one pixel can be better than that of the CCD pixel size.
Since the position uncertainty for corner events is much better than the
pixel size, we can improve the Chandra ACIS spatial resolution by
selecting only these events.
 
We have analyzed X-ray images obtained with the Chandra ACIS for six
point-like sources observed near the optical axis.  The image quality
near the optical axis is characterized by a half power diameter (HPD) of
$0.^{\!\!\prime \prime}66$ that is a convolution of the PSF of the HRMA
and the CCD pixel shape ($24\,\mu$m square).  By considering only corner
events the image quality is improved to $0.^{\!\!\prime \prime}56$
(HPD), which is very close to the image quality of the HRMA alone.  We
estimated the degradation of the image quality obtained by using all
events, compared to that obtained using only corner events, to be
$0.^{\!\!\prime \prime}33$, which coincides with that expected from the
pixel size.  Since the fraction of the corner events is relatively
small, this technique requires correspondingly longer exposure time to
achieve good statistics.

\end{abstract}


\section{INTRODUCTION}

Chandra is the first X-ray imaging satellite to have an image quality
that is comparable to that of optical images.  The point spread function
(PSF) of the high resolution mirror assembly (HRMA) has half-power
diameter (HPD) of about one-half arcsecond (Chandra Observatory Guide),
which corresponds to $24\,\mu$m on the focal plane.  The detector system
should oversample the PSF in order to achieve the highest imaging
capability.  Chandra has two types of non-dispersive detector systems:
the Advanced CCD Imaging Spectrometer (ACIS), a CCD array, and the High
Resolution Camera (HRC), a micro-channel plate.  The CCD chip employed
in the ACIS is a frame-transfer device with $24\,\mu$m square pixels
(Burke et al. 1997) and moderate energy resolution. The overall PSF of
the Chandra/ACIS image is a convolution of the PSF of the HRMA and the
pixel shape of the CCD.  Because the pixel size of the CCD is comparable
to the HPD of the HRMA, the image quality is somewhat degraded. The HRC
has higher spatial resolution ($\sim 20 \mu$m FWHM, with $6.4 \mu$m
pixels), but with poor energy resolution. The observer therefore needs
to choose the detector system depending on the required combination of
angular resolution and energy resolution.

Because the CCD pixel boundaries are sharply delineated by physical
structures on the detector, we can utilize the morphology of X-ray
events that split charge between pixels to provide detector position
information on a size scale much smaller than the actual pixel size.  In
\S\ref{sec:charge}, we explain the process of charge generation and
division within the CCD pixel.  In \S\ref{sec:method}, we describe our
method for subpixel resolution. In \S\ref{sec:results}, we apply this
method to Chandra/ACIS data for six point sources, and show that the
method does indeed recover the inherent angular resolution of the HRMA,
although at the price of reduced observing efficiency.

\section{X-RAY DETECTION IN A CCD}
\label{sec:charge}

The process of detection of X-rays in a CCD has been discussed in detail
by Lumb \& Nousek (1993) and by Townsley et al. (2001).  Here we provide
an abbreviated discussion of these processes sufficient to explain our
method for subpixel resolution.  A CCD is ordinarily used in
photon-counting mode to detect X-rays. To make this possible, the count
rate must be low enough (and the exposure short enough) so that at most
one photon is detected per frame in any given 3 x 3 ``neighborhood'' of
pixels. When an X-ray photon enters the CCD, it is captured by
photoelectric absorption, and generates a primary charge cloud in which
the number of electrons is proportional to the incident X-ray energy.
An X-ray photon of 1\,keV, for example, generates about 275 electrons,
while an optical photon generates only one or two electrons.  
Figure \ref{schematic} schematically shows the evolution of a primary
charge cloud generated by X-ray photons inside the CCD.  The primary
charge cloud expands through diffusion as it travels through the
depletion region of the CCD to the buried channel, where the charge is
collected into pixel structures defined by electric fields near the
surface of the CCD. Before being captured into the photon entrance
pixel, the primary charge cloud generated by an X-ray photon absorbed
near the back of the depletion region can spread into neighboring
pixels, creating a \lq split-pixel event\rq. By contrast, an X-ray
photon absorbed in or near the buried channel will typically deposit its
charge in a single pixel, and the resulting charge packet is referred to
as a `single pixel event'. The incident X-ray energy is estimated by
summing up the signal contained inside the event.  Because the output
from each pixel contains noise, single-pixel events usually have a
better energy resolution than do split pixel events.

The size of the primary charge cloud depends on the travel distance in
the depletion layer and the electric field inside the CCD. If the CCD is
front-illuminated (FI), the travel distance in the depletion region
depends on where the photoabsorption occurs, and therefore depends
strongly on the photon energy. The shorter the attenuation length for an
X-ray photon, the smaller the primary charge cloud will be. The X-ray
photons used in our analysis are mainly around 1\,keV (near the peak
effective area of the instrument), and have a relatively short
attenuation length in silicon compared with the depletion depth or the
pixel size of the CCD. Therefore, they are photoabsorbed in a relatively
shallow region of the depletion layer. On the contrary, if the CCD is
back-illuminated (BI), a 1 keV photon is absorbed near the back surface;
the travel distance will be the depth of the depletion layer, and is
almost independent of the X-ray energy. Applying a simple diffusion
model, the primary charge cloud size will be a few $\mu$m for the BI CCD
and smaller for the FI CCD.  This explains qualitatively why objects
observed with the FI CCD show a higher fraction of single-pixel events
in the branching ratio than those observed with the BI CCD.

The spreading of the charge cloud is less than typical CCD pixel sizes,
and X-ray events 
may contain charge in more than one adjacent pixel in excess of the
local bias level and/or dark current, with most true X-ray events (as
opposed to particle-induced background events) containing no more than
four adjacent pixels. X-ray events are classified by their morphology,
with two common systems being the 8 grades used by the ASCA satellite
(grades 0$-$7) and the 256 grades used by the ACIS instrument (fltgrades
0$-$255).  Further details regarding ASCA grades are given in Yamashita
et al. (1997) while \S 6.4 of the Chandra Proposers' Observatory Guide
(2000) describes ACIS fltgrade.

\section{SUBPIXEL RESOLUTION METHOD FOR CHANDRA/ACIS}
\label{sec:method}

Because the primary charge cloud generated inside the CCD expands
through diffusion, the position on the CCD where the X-ray entered can
be precisely determined if the center of gravity of this charge cloud
can be measured. If the primary charge cloud remains within a single
pixel (a single pixel event), its centroid cannot be determined, and one
generally assumes that the X-ray photon landed in the center of that
pixel, with an uncertainty of about 1/2 pixel. However, if an X-ray
photon is absorbed near a pixel boundary, the primary charge cloud
splits into the adjacent pixel(s), forming a split-pixel event, and the
centroid of the charge distribution can be used to precisely locate, in
at least one dimension, the position at which the photon was
absorbed. The charge cloud size determines the size of the boundary
region within which the X-ray photon becomes a split event.  The pixel
is therefore divided into three classes of regions that characterize
single-pixel events (from the center of the pixel), two-pixel split
events (from the pixel edges) and three- or four-pixel split events
(from the pixel corners).  The ratios among the areas for these three
regions can be roughly estimated from the branching ratio of these X-ray
event grades.

This technique has been verified by our mesh experiments (Tsunemi,
Yoshita, \& Kitamoto 1997; Tsunemi et al. 1998).  In these experiments,
we placed a metal mesh with many small periodically-spaced holes just
above the CCD. Such a setup enables us to directly measure the primary
charge cloud shape (Hiraga et al. 1998).  We thus found that the charge
cloud shape is well expressed by a Gaussian function with a standard
deviation ($\sigma$) for $1\sim4$\,keV photons of about $1\sim2\,\mu$m
(Tsunemi et al. 1999).  Although the charge cloud shape depends on the
CCD chip, the charge cloud size for an X-ray photon with energy of a few
keV is at most a few $\mu$m.  Knowledge of the grade of an event,
combined with the relatively narrow size of the charge cloud, allows the
position at which the X-ray enters the CCD to be determined with
subpixel resolution (Hiraga, Tsunemi, \& Miyata 2001).

Based on these results, we can safely estimate the primary charge cloud
size generated by an X-ray photons with energy of a few keV inside the
ACIS CCD chip to be a few $\mu$m.  The mesh experiment applied to the
ACIS CCD shows that split events are generated very near to the pixel
boundary (Pivovaroff et al. 1998). The grade information of the X-ray
event will thus specify the entering position within the pixel to high
accuracy for X-rays absorbed near the pixel boundaries.  For
single-pixel events, the photon can only be localized to a region
slightly smaller than the pixel size. The two-pixel split events can be
identified only with a pixel edge, which provides high resolution in one
dimension.  The greatest improvement for the X-ray position resolution
is achieved for three- or four-pixel split events (corner events), which
are generated when the X-ray enters the CCD very close to a pixel
corner.  The X-ray entering position must be within one charge cloud
size from the pixel corner.  This allows us to improve on the spatial
resolution of the pixel size by roughly a factor of ten, based solely on
the event grade.

This is of practical use for data obtained by Chandra
because, unlike the HST, 
which points very accurately at a fixed position during
observations, Chandra is intentionally moved slowly across
the sky during each observation. This `dithering' motion of the
observatory moves the target image across the CCD surface, forming a
Lissajous figure with an amplitude of about 20 pixels and a period of
about 1000 sec.  Therefore, X-rays from a point source do not always
enter the same subpixel position of the same pixel, but have randomly
distributed subpixel positions.  Some photons enter the central
part of the pixel, forming single-pixel events, while others enter 
the pixel boundary, forming split pixel events. The latter can be used
to achieve higher spatial resolution than that determined by the pixel
size.  
Strictly speaking, the image shape can vary according to the dithering
motion.  With taking into account the amplitude of the dithering motion,
the degradation of the image quality concerned is negligibly small
compared to the actual PSF of the HRMA.

The method we employ is quite simple.  First, we discard all X-ray
events except for corner events (ACIS fltgrades 10, 11, 18, 22, 72, 80,
104, and 208).  We know that these remaining events entered the detector
near a pixel corner, but were projected onto the sky as if they had
entered at the pixel center.  We therefore need to shift the sky
coordinates for these events by 1/2 pixel along each axis in {\em
detector} coordinates. To do this, we need to know the orientation of
the CCD chip projected onto the sky coordinates. This is given by the
roll angle of the spacecraft, which is specified in the header keyword
of the event file.  We then can shift the position for corner events by
half the pixel size (which is $0.^{\!\!\prime\prime}246$) along both
pixel sides.  The actual shift direction depends on the CCD chip
employed and the roll angle as well as the fltgrade of the event. The
resulting sky positions accurately reflect the actual incident positions
on the CCD detector to within about $0.^{\!\!\prime\prime}05$ for
perfect attitude knowledge.

\section{RESULTS}
\label{sec:results}

To test this method, we searched our GTO data and publicly available
calibration data for point-like sources with appropriate
characteristics.  We selected sources with good statistics (at least 1500
total counts) and that are close to the optical axis (within
75$^{\prime\prime}$) in order to avoid possible distortion. We selected
six point-like sources; two stars and four AGNs. The two stars, which we
refer to as \lq source 1\rq\ (${\rm 5^h 35^m 15.^{\!\!s}}67$, $-5^\circ
23^\prime 11.^{\!\!\prime\prime}2$) and \lq source 2\rq\ (${\rm 5^h 35^m
15.^{\!\!s}}73$, $-5^\circ 23^\prime 15.^{\!\!\prime\prime}5$), are in
the Orion Nebula.  Two AGNs, \AGNa\ and \AGNb, come from the public
data.  The other source contains the two gravitationally lensed images
of the quasar \AGNL, named \lq image A\rq\ and \lq image B\rq\ (Walsh,
Carswell, \& Weymann 1979). Three sources are observed with the
front-illuminated (FI) CCD while the others are observed with the
back-illuminated (BI) CCD.

We removed the subpixel randomization usually applied in the standard
data processing at the Chandra X-ray Center so that we can assess the
precise X-ray entering position on the detector.  Townsley et al. (2000)
developed a CTI correction method that could correct the `grade
migration' (the corruption of event grades caused by poor charge
transfer efficiency in the FI CCDs) for the data obtained after 1999
September 16.  We applied their method to five data sets out of six,
since it could not cover the data for \AGNa.  The observational
configurations are summarized in Table~\ref{table:summary}.

We then sorted all the X-ray events (all events) according to the event
geometry on the CCD.  Specifically, they were sorted as single-pixel
events, two-pixel split events and corner events.  Furthermore, they
were sorted according to the event geometry within the $3\times 3$
pixels using the \lq fltgrade\rq\ (see the Chandra Proposers'
Observatory Guide).  In this way, we sorted the event list into nine
groups: one single-pixel event group, four two-pixel split event groups
and four corner event groups. The statistics for each grade are
summarized in Table~\ref{table:summary}.  We then extracted the images
in absolute sky coordinates, which places each event on the sky at the
projection of the central pixel of the $3\times 3$ pixel event
neighborhood.

Figure~\ref{event_dist} shows an example of the distributions of four
groups of corner events for \AGNL\ image A in absolute sky
coordinates 
that is obtained by employing uncorrected events.  
One can clearly see the systematic shift of their centers
of gravity, depending on the fltgrade of the
event. Figure~\ref{CG_event} shows the centers of gravity for all nine
event types discussed here in absolute sky coordinates for six sources,
where the square shows the pixel shape projected on the sky with the
proper orientation.  
These figures are also obtained by using uncorrected events.  
We confirm that the distribution of the single-pixel
events coincides with the nominal position, which is the center of the
central pixel of the event. As expected, the centers of gravity of the
two-pixel events lie near the centers of the pixel edges, while the
centers of gravity of the three- and four-pixel events fall near the
pixel corners.
Because the software places these uncorrected events at the centers of
the pixels instead of at their true corner locations, they end up at
incorrect celestial locations with the centroids of the four corner
event groups making an inverted image of the pixel.  Applying the
correction discussed in this paper would make the four pixel clouds lie
on top of one another.

Figure~\ref{image} shows images obtained using corrected corner events, as
well as those obtained from all events for comparison.  Since we find
that all images are almost point symmetric, we fit them with a
point-symmetric Gaussian function with constant background.  The
background is less than 1\% of the count rate of the source.  We then
calculated the HPD 
from the best fit Gaussian function for each image. The results are
summarized in Table~\ref{table:image_size} where all errors are 90\%
confidence levels. HPD$_{all}$ and HPD${_c}$ represent the HPDs for the
images for all events and for the corner events, respectively.

The actual HPD can be easily measured by counting the number of
events. In this measurement, we assume the center of the Gaussian
function determines the image center.  We then count the number of
events contained in a circle as a function of diameter.  We define the
actual HPD as the diameter of the circle that contains half of the
events appeared in a circle of $4^{\prime\prime}$ in diameter.  We
confirmed that the actual HPD measured are consistent with those
obtained by fitting the Gaussian function within statistical
uncertainties.

\section{DISCUSSION}

The PSF of the HRMA has a size of about $0.^{\!\!\prime \prime}5$ HPD on
the optical axis, which corresponds to about $24\,\mu$m on the detector.
The HPD of the final image is the root mean square of that for the HRMA
and that for the CCD pixel.  The $\sigma$ for a square distribution is
$\sqrt{a^2/12}$ where \lq a\rq\ is the square size.  If we use
$2\sqrt{2\ln 2}$ as the conversion factor from $\sigma$ to HPD just as
the case of the Gaussian function, the HPD for the ACIS CCD pixel is
about $16\,\mu$m which is the HPD for all events if the subpixel position
of the photon is unknown.

Therefore, the HPD of the final image obtained for all events will be
$29\,\mu$m on the detector or $0.^{\!\!\prime \prime}59$ when projected
onto the sky.  However, the position accuracy of the corner events is
about a few $\mu$m since their distribution is within the charge cloud
shape. This value is negligibly small compared with the PSF of the HRMA,
and does not degrade it.

We can estimate the improvement of the position resolution between all
events and the corner events by $\Delta {\rm HPD} = \sqrt{{\rm
HPD}_{all}^2 - {\rm HPD}_c^2}$, as listed in
Table~\ref{table:image_size}. Figure~\ref{delta_sigma} shows a plot of
$\Delta {\rm HPD}$ along the angular distance from the optical axis as
well as ${\rm HPD}_{all}$ and ${\rm HPD}_c$.  We found $\Delta {\rm HPD}
= 0.^{\!\!\prime \prime}35\pm0.^{\!\!\prime \prime}07$ by using all the
data. This corresponds to $17\pm 3\,\mu$m on the detector.  This shows
that by using only corner events we can significantly reduce the PSF
degradation caused by the under-sampling of the PSF by the ACIS
pixels. Practically, we can say that the image degradation by all events
is noticeable while that by the corner events is negligibly small.

In principle, the effective resolution could be improved further by
using both the knowledge of the charge cloud shape and that of the grade
of the multi-pixel events to determine the exact position of the X-ray
within the boundary region.  Actually the primary charge cloud shape
{\it is} known an the position on the CCD where the X-ray entered {\it
can} be determined for split pixel events with sub-$\mu$m resolution
(Hiraga et al. 2001).  The simplifying assumption that we have made here
is that, even if the charge cloud shape is not known, we can assume that
the X-ray entering position for corner events is precisely at the pixel
corner. For Chandra, the angular uncertainty associated with this
assumption is much smaller than the PSF of the HRMA. These data
demonstrate that the boundary region itself is sufficiently small that
this additional information would not significantly improve the PSF.

The fraction of the corner events, which depends on the incident X-ray
spectrum as well as on the CCD type, is about $4\sim 16\,\%$ of the
total events. Thus, this technique, while able to improve the effective
angular resolution of Chandra, does so at the cost of low
efficiency. However, we have shown that this technique is effective,
even for as few as 300 corner events.

\section{CONCLUSION}

We have demonstrated that X-ray split events on the CCD can yield
position information more precise than the pixel size. The Chandra
Observatory has a dither motion that moves the target on the detector so
that the X-rays are absorbed uniformly over the pixels.  Therefore, some
fraction of the X-ray photons always generate corner events that can be
used to achieve the finer image without degradation by the CCD pixel
size.

We have studied six point-like sources that were observed near the
optical axis.  Since the split events are generated near the CCD pixel
boundary, we can determine the X-ray entering position with much better
resolution than the pixel size.  The average image size using the corner
events is $0.^{\!\!\prime \prime}56$ (HPD) while that using all events
is $0.^{\!\!\prime \prime}66$ (HPD).  The image degradation between the
image obtained by the corner events and the image obtained by all events
corresponds to that expected from the CCD pixel shape. Therefore, we can
say that the image obtained using the corner events is almost free from
the degradation of the image due to the CCD pixel shape. The corner
events are generated only when X-rays enter near the pixel corner, which
reduces the effective count rate.  We therefore need relatively longer
exposure times to obtain good statistics.  
Finally we should add that our method is useful only for the Chandra
HRMA that is undersampled by the detector. Thus, for example, this
technique cannot be applied to XMM-Newton observations.

The authors would like to express their special thanks to Dr. Eric
Feigelson who kindly supplied us his Chandra data set and
Dr. M. W. Bautz of MIT for useful discussions.  Dr. Jonathan McDowell
explained to H.T. the attitude determination system of Chandra in
detail.  This research is partially supported by ACT-JST Program, Japan
Science and Technology Corporation, the Sumitomo foundation, and by NASA
contract NAS8-38252.

\clearpage

\newcommand{\Lower}[1]{\smash{\lower 2ex \hbox{#1}}}

\begin{deluxetable}{lccccccc}
\tabletypesize{\scriptsize}
\tablecaption{{\sc Summary of the Observations} \label{table:summary}}
\tablewidth{0pt}
\tablehead{
\Lower{ }	& & \multicolumn{2}{c}{Orion Nebula} &
\Lower{PG1634+706} &\Lower{PKS0312-770}& \multicolumn{2}{c}{Q0957+561}\\
 & & \colhead{source 1} & \colhead{source 2}& & & \colhead{image B} &
\colhead{image A}
}
\startdata
Obs ID		&& 18	   & 18    & 69   & 1109  & 362   & 362 \\
Category	&& Star    & Star  & AGN  & AGN   & AGN   & AGN \\
CCD chip	&& FI	   & FI    & BI   & FI	  & BI	  & BI \\
Exposure (ks)	&& 46.94   & 46.94 & 4.80 & 13.00 & 47.26 & 47.26 \\
Distance (arcsec)\tablenotemark{a}
		&& 6.9	   & 11.   & 19.  & 20.   & 69.   & 72. \\
 CTI correction
		&& Yes	   & Yes   & Yes  & No	  & Yes   & Yes \\
 \hline
 Event grade		& fltgrade & \multicolumn{6}{c}{Number of events}\\
 \hline
 Single-pixel		&0	& 5663& 5132& 607 & 1321& 4303 & 5763 \\
 2-pixel split		&	&     &     &	  &	&      &	\\
 \hspace{2ex}down	&2	& 364 & 279 & 241 & 238 & 1135 & 1478	\\
 \hspace{2ex}left	&8	& 683 & 383 & 198 &199	& 1084 & 1429	\\
 \hspace{2ex}right	&16	& 685 & 348 & 232 & 188 & 1078 & 1408	\\
 \hspace{2ex}up 	&64	& 864 & 531 & 223 & 393 & 1125 & 1469	\\
 Corner 		&	&     &     &	  &	&      &	\\
 \hspace{2ex}left-down	&10,11	& 220 & 72 & 83 & 70	& 328 & 475	\\
 \hspace{2ex}right-down &18,22	& 227 & 65 & 65 & 74	& 329 & 456 \\
 \hspace{2ex}left-up	&72,104 & 202 & 82 & 58 & 104	& 305 & 392	\\
 \hspace{2ex}right-up	&80,208 & 215 & 67 & 73 & 96	& 308 & 474	\\
\enddata
\tablenotetext{a}{Angular distance from the optical axis in arcseconds}
\end{deluxetable}

\clearpage

\begin{deluxetable}{lcccccc}
\tabletypesize{\scriptsize}
\tablecaption{{\sc Image size (HPD [arcseconds])} \label{table:image_size}}
\tablewidth{0pt}
\tablehead{
\Lower{target}	& \multicolumn{2}{c}{Orion Nebula} &\Lower{PG1634+706}	&
\Lower{PKS0312-770}
& \multicolumn{2}{c}{Q0957+561}\\
 & \colhead{source 1}	& \colhead{source 2}& & & \colhead{image B}	&
\colhead{image A}
}
\startdata
HPD$_c$  & $0.52\pm 0.02$ & $0.60\pm 0.04$ & $0.61\pm 0.04$
   & $0.49\pm 0.02$ & $0.67\pm 0.02$ & $0.67\pm 0.02$	\\

HPD$_{all}$ & $0.64\pm 0.01$ & $0.71\pm 0.01$ & $0.69\pm 0.02$
   & $0.65\pm 0.02$ & $0.74\pm 0.01$ & $0.75\pm 0.01$	\\

$\Delta {\rm HPD}$\tablenotemark{a}

  & $0.37\pm 0.03$ & $0.37\pm 0.06$ & $0.32\pm 0.09$
   & $0.43\pm 0.04$ & $0.33\pm 0.05$ & $0.33\pm 0.04$	\\
\enddata
\tablenotetext{a}{$\Delta {\rm HPD} = \sqrt{{\rm HPD}_{all}^2 - {\rm
HPD}_c^2}$.}
\tablecomments{Errors quoted are 90\% confidence levels.}
\end{deluxetable}

\clearpage

\begin{figure}

\centerline{FIGURE CAPTIONS}

\caption{Schematic view of the evolution of a primary charge cloud
generated by X-ray photons inside a front-illuminated CCD}\label{schematic}

\caption{Plot of the distribution of corner events for \AGNL\ versus
absolute sky coordinates.  The four groups, left-down (black),
right-down (red), left-up (green) and right-up (blue), are clearly
separated from each other. The square represents the ACIS pixel in its
proper orientation.}  \label{event_dist}

\caption{Plot of the distribution of the centers of gravity for corner
events (red), two-pixel split events (blue) and single pixel events
(black) versus the absolute sky coordinates for the Orion Nebula (a)
source 1, (b) source 2, (c) \AGNa, (d) \AGNb, \AGNL\ (e) image A and (f)
image B.  The uncertainties for the centers of gravity are comparable to
or smaller than these marks.  The square shows a pixel shape in its
proper orientation.} \label{CG_event}

\caption{X-ray images for six sources: the Orion Nebula (a) source 1,
(b) source 2, (c) \AGNa, (d) \AGNb, \AGNL\ (e) image A and (f) image
B. The left side of each figure is the image obtained with corner events
shifted by half pixel size while the right side is obtained with all
events.  Contour levels are on a  linear scale, showing 20\%, 40\%, 60\%,
and 80\% of the peak value.}  \label{image}

\caption{Plot of the HPD of image sizes obtained with all events and corner
events as a function of the angular distance from the optical
axis. $\Delta {\rm HPD}$ is also plotted.  } \label{delta_sigma}
\end{figure}

\end{document}